\documentclass[twocolumn,fleqn]{svjour2}    % twocolumn
\smartqed  % flush right qed marks, e.g. at end of proof
\usepackage{graphicx}
\usepackage{subfigure} % use for side-by-side figures

\newcommand \be{\begin{eqnarray}}
\newcommand \ee{\end{eqnarray}}
\newcommand \bee{\begin{equation}}
\newcommand \eee{\end{equation}}

\journalname{EPJE}
\begin{document}

\title{Formation of brine channels in sea-ice%\thanks{Grants or other notes
}


\author{Klaus Morawetz $^{1,2,3}$     \and Silke Thoms
$^4$    \and Bernd Kutschan$^5$         %etc.
}


\institute{\at $^1$ 
              M\"unster University of Applied Sciences, Stegerwaldstra\ss{}e 39, 48565 Steinfurt, Germany\\
\at $^2$ 
International Institute of Physics (IIP), 
%Federal University of Rio Grande do Norte, 
Av. Odilon Gomes de Lima 1722, 59078-400 Natal, Brazil\\
\at $^3$
Max-Planck-Institute for the Physics of Complex Systems, 01187 Dresden, Germany\\
 \at $^4$ Alfred Wegener Institut, Am Handelshafen 12, D-27570 Bremerhaven, Germany
%              \email{morawetz@fh-muenster.de}           %  \\
\\
\at $^5$ 
              M\"unster University of Applied Sciences, Corrensstra\ss{}e 25, 48149 M\"unster, Germany\\
}

\date{Received: date / Accepted: date}

\maketitle

\begin{abstract}
Liquid salty micro-channels (brine) between growing ice platelets in sea ice are an important habitat for $CO_2$ - binding microalgaea with great impact on polar ecosystems.  The structure formation of ice platelets is microscopically described and a phase field model is developed. The pattern formation during
solidification of the two-dimensional interstitial liquid is considered by two coupled order parameters, the 
tetrahedricity as structure of ice and the salinity. The coupling and time-evolution of these order parameters are described by a consistent set of three model parameters. They determine the 
velocity of the freezing process and the structure formation, the phase diagram, the super-cooling and super-heating region, and the
specific heat. The model is used to calculate the short-time frozen micro-structures. The obtained morphological structure is compared with the vertical brine pore space obtained from Xray computed tomography. 
\end{abstract}

\section{Introduction}
Sea-ice does not freeze homogeneously but some liquid salty micro-channels remain which are called brine. These brine capillaries are an important habitat for $CO_2$ - binding microalgaea with great impact on the polar ecosystems.
Their carbon consumption amounts to
about 18\% of the entire carbon consumption in the southern ocean.
Therefore it is desirable to understand the formation of such brine channels as one possible habitat for carbon-binding algae.
Two-phase regions of pure ice crystals and water are also known as mushy
layers in the context of binary alloys \cite{WW97,FUWW06}.
  Highest cell abundances occur in these regions, due to the higher porosity and
due to the constant flushing with nutrient-rich seawater \cite{Ackl,wer}.

The freezing process of salty water is one example of the
solidification of binary alloys \cite{Till,Cha}. Models of ice polluted with any salt as ”'liquid jelly”' \cite{Qui2} consider this process as first-order phase transitions \cite{B87}. Sometimes, for solidification of seawater, the model of percolation transitions is used in brine trapping  \cite{Gol,GOl1,Gol2}.
 In this respect a morphological stability analysis was applied 
to the solidification of salty water \cite{W92}. 
  All these quantitative models  \cite{Co1,Co2,Co3} have investigated the brine channel volume, salinity profile or heat expansion, but have unfortunately not considered the pattern formation. Here we will present a dynamical model exploring the formation of morphological patterns consistent with the thermodynamics of freezing. Concentrating on the short-time evolution we consider the structure-forming processes here as adiabatic and neglect the heat transport.

Images of single crystals in sea ice with the help of
X-ray computed tomography \cite{Pringle} show arrays of nearly parallel brine layers whose connectivity and complex
morphology varies with temperature. The pore space turns out to be much more complicated than
suggested by simple models of parallel ice lamellae and parallel brine sheets \cite{weeks}.  Some\-times the granular sea ice text\-ure is imagined to arise from a deposition of fragile ice crystals. They are thought to be formed within the turbulent ocean interior and then rising buoyantly to the ocean surface \cite{J94,PeEi}. In these settings the size of the settling crystals plays a dominant role in controlling the observed structures. We consider here the opposite view that these structures result from a thermodynamic instability during growth itself rather than from the external deposition.

In order to describe a realistic pattern formation and the phase transition on the same theoretical basis we use a phase-field model for the solidification of the two - dimensional interstitial liquid. We will calculate the frozen micro-structures and will compare with the vertical brine pore space obtained from X-ray computed tomography \cite{Pringle,Wei}.
The aim is to present a model with the smallest possible number of
microscopic parameters to be extracted from experiments. We find here that three parameters are sufficient, the freezing, the
structure, and the diffusivity parameter. Only the first two ones determine
the phase diagram while the diffusivity enters the brine channel size. 
The linear stability analysis leads then to the parameter
range where structure can appear and the numerical solution will allow to
compare with the experimental data.

The outline of the paper is as follows. First we develop the minimal model and give the meaning of different used model parameters. Then we derive the thermodynamics of supercooling and freezing point depression providing the phase diagram. In chapter IV we will discuss the linear stability analysis which yields the most unstable modes and scales. Then we determine the model parameters from the properties of water in chapter V. The time evolution is presented by a numerical solution of the coupled phase-field model in chapter VI and is compared with the experiment in chapter VII. Chapter VIII summarizes and discusses shortcomings as suggestions for further investigations.

\section{Phase-field model}
%\section{Model development}

%\subsection{Structure properties and order parameter of water and ice}

To distinguish between ice- and
water molecules we use a two-state function,
the ''tetrahedricity' \cite{Me}  
\be
u =1-\frac{1}{15 <l^2>}\sum_{i,j}(l_i - l_j)^2 ,
\ee
where the $l_i$s are the differences of the six edges of the tetrahedron
formed by the four nearest neighbors of the considered water molecule. For an
ideal tetrahedron one has $u=1$ and the random structure is represented by
$u=0$.  We assume the standard expansion of the energy function in powers of this order parameter \cite{B87,Beste}
\be
%\!\!\!\!\!\!\!\!
\frac{D_{\rm ice}}{2}(\nabla u )^2 \!+\!
  \frac{a_1}{2}u^2 \!-\!  \frac{a_2}{3}u^3 \!+\!
  \frac{a_3}{4}u^4 \! +\! \frac{h}{2}u^2v \!+\! \frac{D_{\rm salt}}{2}v^2.
\label{free_energy}
\ee
Here we have coupled additionally a second order parameter, the salinity $v$, by the term $h$ which can be
considered as reaction rate between water and ice.
The parameter $a_1$
is the freezing parameter determining the phase transition, the structure
parameter $a_3$ is responsible for nonlinear behavior and $D_{\rm ice}$ and
$D_{\rm salt}$ are the
diffusion coefficients of ice and salt. The coefficient $a_2$ is connected
with an uneven exponent and is therefore responsible for the phase
transition of first kind.   All these parameters depend
on the temperature and can be scaled to
only three relevant parameters. The phase diagram will be determined only by
two of them, the dimensionless structure and freezing parameter.

The coupling of the two order parameters is chosen in a form which enables the
conservation of the total mass of the salt as follows.
We demand a balance equation of the form $\partial v/\partial t=-\nabla \vec j$
where the current is assumed to be proportional to a generalized force $\vec
j\sim \vec F$ which should be given in terms of a potential $\vec F=-\nabla
P$. This potential in turn is expressed by the variation of the free energy density 
$P=\delta {f}/\delta v$. This procedure is nothing but the second law
of Fick and we obtain an equation of the
Cahn-Hilliard-type without the fourth derivation for the evolution of
the salinity $v$. 
%\cite{Thoms}

Defining the reduced time $
\tau={D_{\rm salt} a_2^2t}/{h^2}$, the spatial coordinates 
$\xi={a_2x}/{h}$, the dimensionless order parameters of ice/water structure $\psi={h^2u}/{D_{\rm salt} a_2}$, and the salinity
$\rho={h^3v}/{D_{\rm salt} a_2^2}$,
we obtain the coupled order-parameter equations
\begin{eqnarray}
\frac{\partial \psi}{\partial \tau} &=& -\alpha_1'\psi + \psi^2 -
\alpha_3\psi^3 - \psi\rho + D\frac{\partial^2 \psi }{\partial \xi^2} 
\nonumber\\
\frac{\partial \rho}{\partial \tau} &=& 
\frac{1}{2}\frac{\partial^2 \psi^2}{\partial \xi^2} + 
\frac{\partial^2 \rho}{\partial \xi^2}. 
\label{f2}
\end{eqnarray}

These time-dependent Ginzburg-Landau
differential equations couple the dynamics of the dimensionless
order parameter $\psi$ and the dimensionless salinity $\rho$ depending
only on three parameters, the freezing parameter
$
\alpha_1'={a_1 h^2}/{a_2^2 D_{\rm salt}}
%\label{a10}
$, 
the structure parameter $\alpha_3={a_3
  D_{\rm salt}}/{h^2}$, and the diffusivity $D={D_{\rm ice}}/{D_{\rm salt}}$  with
$\alpha_1,'\,\alpha_3,\,D > 0$.
The Eq.s (\ref{f2}) represent a modification of the model C in the Hohenberg-Halperin classification \cite{Hohenberg}, there eq. 4.50. The difference here is an additional quadratic term in the first equation coming from the uneven exponent in (\ref{free_energy}) with $a_2$ responsible for the first-order phase transition. We neglect in this model any velocity or temperature field which could be included analogously to the model H in \cite{Hohenberg}. 

\section{Thermodynamics of supercooling, super-heating and freezing point suppression}

The parameters $\alpha_1'$ and $\alpha_3$  describe the regions of
ordered and non-ordered phase. This can be seen from the
uniform stationary free energy density. We therefore use the stationary solution of the second equation in the first one of (\ref{f2}) to obtain
\be
f(\Psi_0,\rho_0) = \frac{\alpha_1}{2}\psi_0^2-\frac{1}{3}\psi_0^3+\frac{\alpha_3}{4}\psi_0^4
\label{4}
\ee
where the temperature-dependent compound
parameters $\alpha_1(T) =  \alpha_1'(T) + \rho_0$ and $\alpha_0  =  \frac{1}{2}\rho_0^2 - \gamma \rho_0$  appear in terms of the 
salinity $\rho_0$. 
Freezing-point
depression occurs since $\alpha_1'+\rho_0$ corresponds to a higher
temperature than $\alpha_1'$. 

The temperature and salinity dependence of $\alpha_3$
is supposed to be weak near the phase transition. At the lower limit of the super-cooling region of
fresh water \cite{Nev,Dor}, $T^0_c = 233.15 K$, the parameters $\alpha_1'$
vanishes linearly for first-order phase transitions \cite{Beste} such that we can assume 
$
\alpha_1'(T)=\tilde{\alpha}_1(T-T^0_c).
$
The freezing point
depression in the framework of Landau-Ginzburg theory can be expressed
therefore as
\be
\Delta T = -
{\rho_0\over
\tilde \alpha_1}
= -
{D_{\rm salt}\over
\tilde a_1}
{a^2_2\over h^2} \rho_0.
\label{dT}
\ee
Introducing the salinity-dependent super-cooling temperature
$
T^0_{c,s} = T^0_{c} - |\Delta T|
$
the freezing parameters $\alpha_1$ depends on the temperature according to
\be
\alpha_1(T) = \rho_0
{T - T^0_{c,s}\over
|\Delta T|}. 
\label{aa0}
\ee

\begin{figure}[h]
(a)\includegraphics[width=8cm,angle=0]{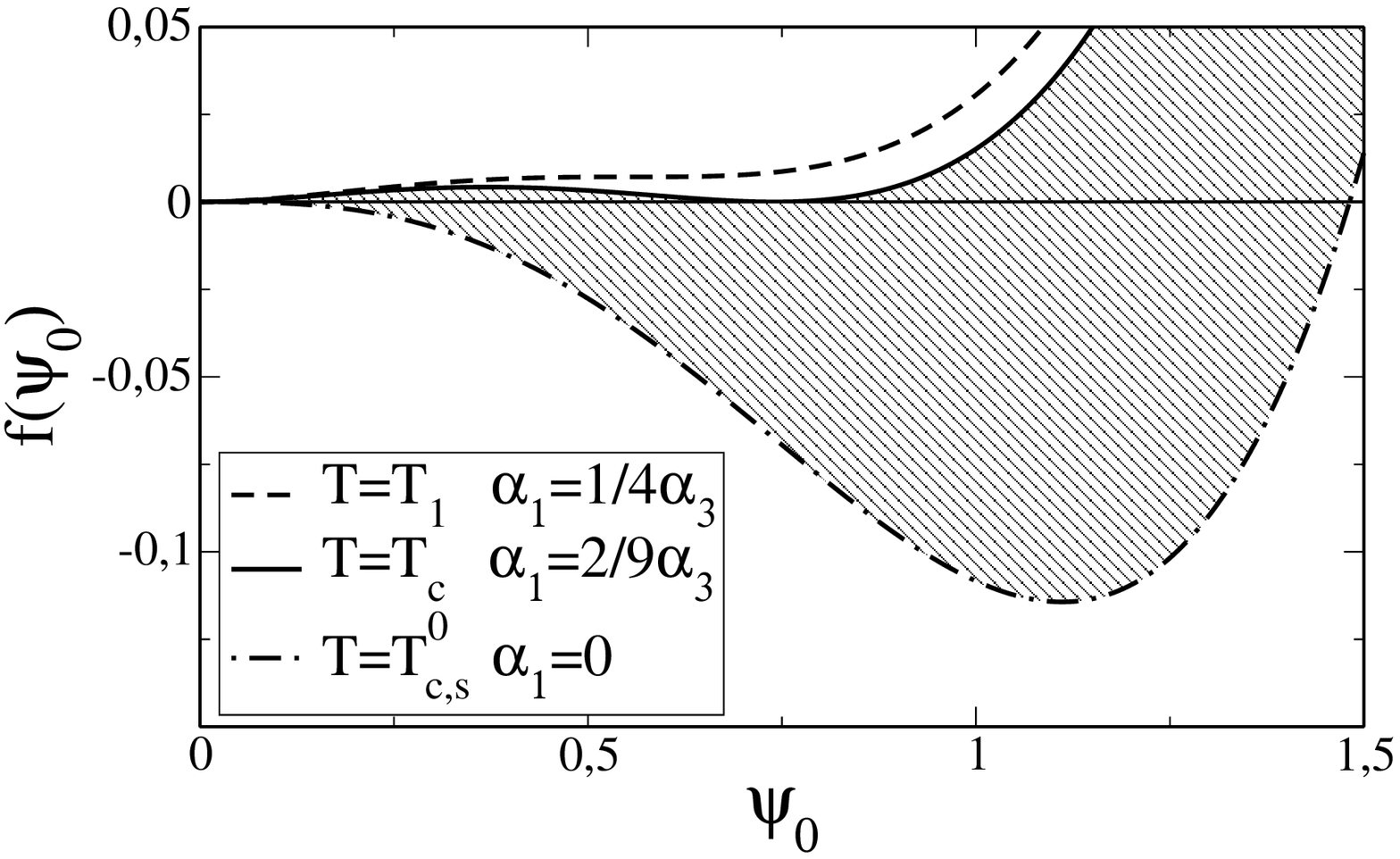}

(b)\includegraphics[width=8cm,angle=0]{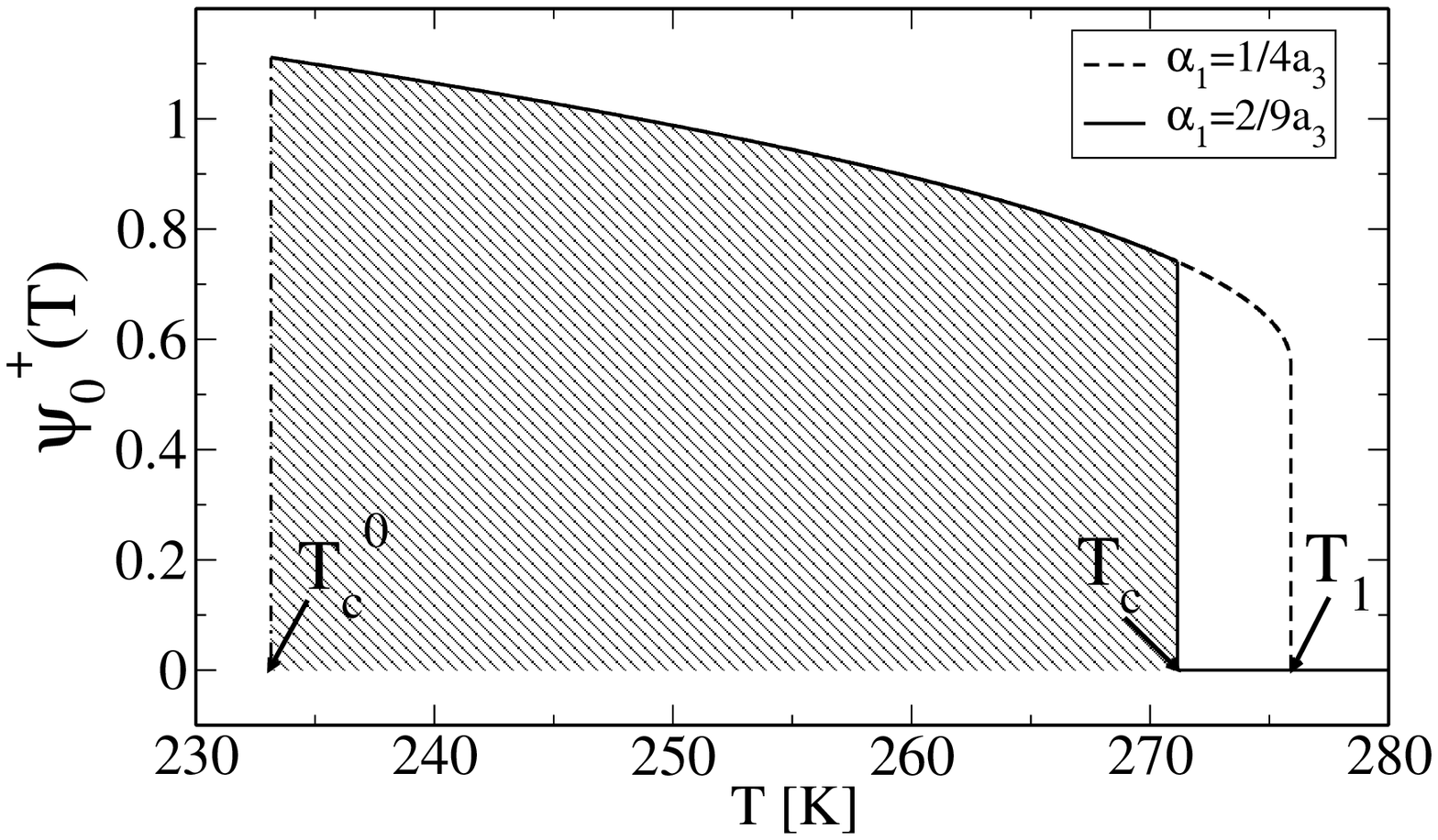}
\caption{Condition for a first-order phase transition. 
(a) The free energy density $f$ versus the uniform dimensionless 
order parameters (tetrahedricity) for some freezing parameters $\alpha_1$ and
the structure parameters $\alpha_3=0.9$ representing the super-cooling region for freezing of
  hexagonal ice $T^0_{c,s}<T_c$ (shaded area) and the super-heating region  $T_c<T<T_1$. 
(b) Dependence of the absolute minimum of the free energy density on $\alpha_1$
(solid line) in temperature-dependent representation. The dashed line corresponds to the position of the second
minimum.
\label{free_energy_temp}}
\end{figure}

The free
energy density (\ref{4}) has a minimum at
$\psi_0^0 = 0$ and a minimum/maximum for
\be
\psi_0^{\pm} = \frac{1}{2\alpha_3}\left (1 \pm
\sqrt{1-4\alpha_1\alpha_3}\right ).
\ee
 For $\alpha_1>1/4\alpha_3$, the minimum at $\psi_0^0 = 0$ is the only 
allowed physical solution, which is the disordered state.  As long as 
\be
\alpha_1 \le {1\over 4\alpha_3}=\alpha_1(T_1)
\label{inst}
\ee 
 a second 
relative minimum appears at $\psi_0^{+}$ as seen in figure \ref{free_energy_temp}a.  The lowest free energy establishes the stable state. The coexistence curve where these two local minims are equal and $f(\Psi_0^+)=f(\Psi_0^0)= 0$ yields the critical temperature 
\be
\alpha_1(T_c)={2\over 9\alpha_3}.
\label{16}
\ee
This coexistence curve is plotted as solid line in Fig.~\ref{free_energy_temp}. 
Above the critical parameters $\alpha_{1}(T_c) < \alpha_1(T) < \frac{1}{4\alpha_3}$ the ordered 
phase $\psi_0^{+} > 0$ is metastable whereas the non-ordered phase ($\psi_0 =
0$) is stable.  For small $\alpha_1\le\alpha_{1}(T_c)$ the second  minimum at $\psi_0^{+}>0$ becomes deeper and the ordered phase $\psi_0^+$ is the stable one. 
Therefore the absolute minimum changes discontinuously from $\psi_0 = 0$ to $\psi_0^{+} > 0$ as plotted in Fig. \ref{free_energy_temp}(b). 
The jump at $T_c$ is a measure for the latent heat
during the first order phase transition between water and ice.

We identify the
upper borderline of a stable structure formation (\ref{16}) with the
freezing temperature since this is the line where structure, i.e. ice
formation is possible at all.  In the same manner the borderline
of metastable structure (\ref{inst}) represents the super-heating temperature. The shaded area in
Fig. \ref{free_energy_temp} describes the super-cooling region between $T_c$ and
$T^0_c$. The latter one is the temperature where $\alpha_1=0$. Above this area we find the super-heating region for $T_c<T<T_1$. From (\ref{inst}) and (\ref{16}) the relation between the super-cooling temperature $T_c^0$, the
freezing temperature $T_c$, and the super-heating temperature $T_1$ reads
\be
T_1=\frac{9}{8}T_c - \frac{1}{8}T^0_c.
\label{45}
\ee

\section{Linear stability analysis}

The linear stability analysis for the two local minim around 
the disordered phase $\psi_0^{0}$ and the ordered phase $\psi_0^{+}$ with
$ \bar \rho = \bar \rho_0  \exp[\lambda(\kappa) \tau +i\kappa\xi]$
leads to the two possible growth rates
\begin{eqnarray}
\lambda_{1,2} = - %\left 
[(D+1)\kappa^2 -{\aleph}  \pm
  \sqrt{\Delta}%\right ]
/2
\label{rootslam}
\end{eqnarray}
with
$
\Delta  =   
 [(D-1)\kappa^2 -{\aleph}]^2 + 4 \kappa^2\psi^2_0 > 0
$
and ${\aleph}=-\alpha_1+2\Psi_0-3 \alpha_3 \Psi_0^2$ which takes the value
${\aleph}=-\alpha_1$ for the fixed point $\Psi_0^0=0$ and ${\aleph}=\psi_0-2\alpha_3\psi^2_0$ for $\Psi_0^\pm$.
Time-oscillating structures would appear only if ${\rm Im}\lambda(\kappa)\neq
0$, i.e. $\Delta<0$, which is not the case in our model.

An unstable fixed point $\lambda(\kappa)>0$ allows any fluctuation 
with a wave-vector $\kappa$ to grow exponentially in time. For the fixed point representing the disordered phase, $\psi_0 = 0$ and $\rho_0=const$,
\be
%\!\!\!\!\!\!\!
\lambda_{1,2}=\frac 1 2\left [ -(D+1) \kappa^2-\alpha_1\pm
  |(D-1)\kappa^2+\alpha_1|\right ]<0
%\nonumber\\
%&&
\ee
and no structure formation occurs in this state which was expected
for the disordered phase, of course.

\begin{figure}[h]
\centerline{\includegraphics[width=8cm]{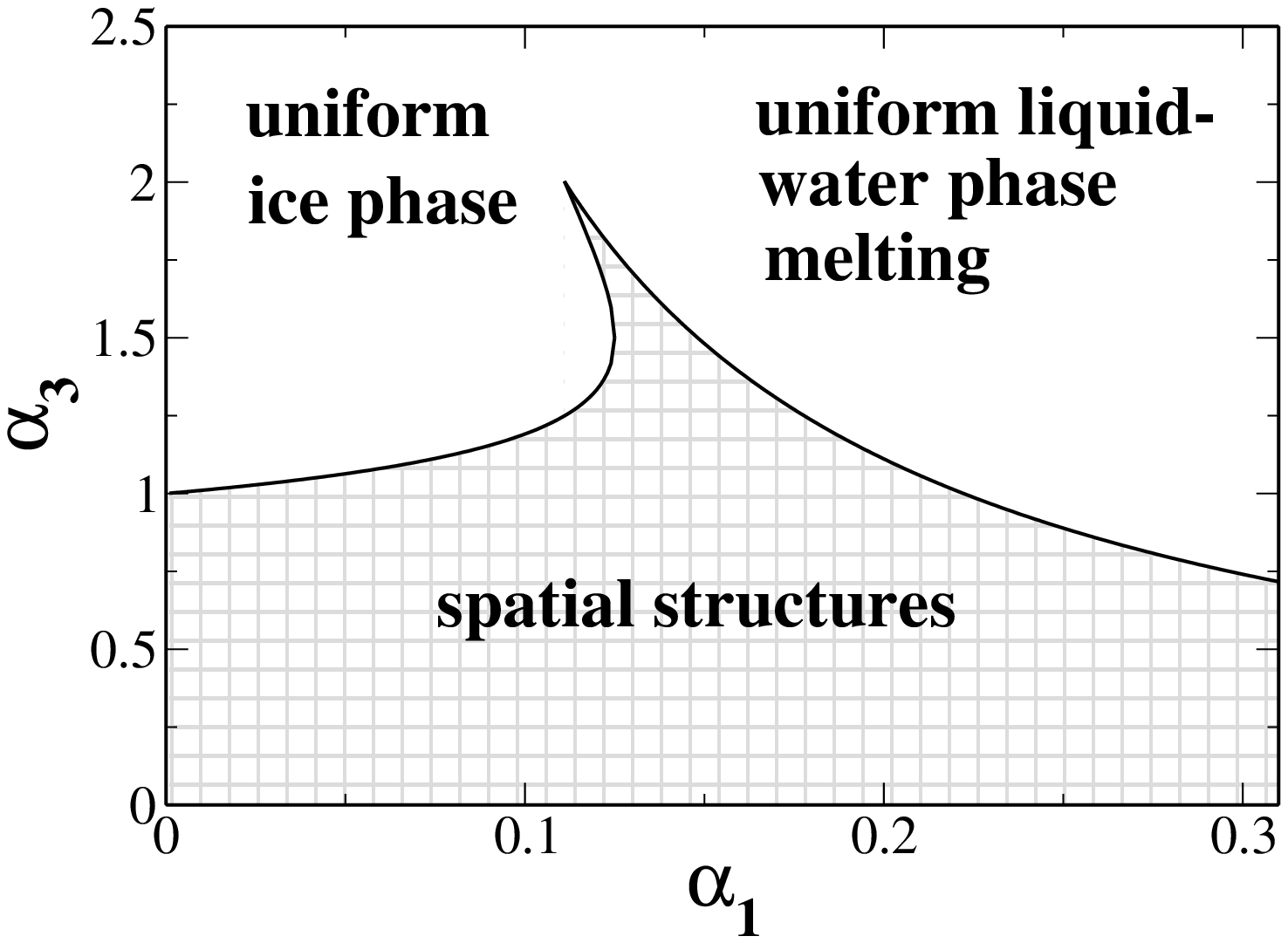}}
\caption{The instability regions of the fixed 
 point $\psi_0^{+}$ and $\rho_0 = const$ as phase
diagram together where spatial structures can occur (checked region).}
\label{InstabReg}
\end{figure}

We can only have 
positive $\lambda(\kappa)$ if the values of $\kappa$ are restricted to the region between the zeros 
of $\lambda(\kappa)$, which is
%\begin{equation}
$ \kappa^2\in %\left
(0,\,{\psi_0^+ } %\left
[1 - (2\alpha_3 - 1)\psi_0^+ %\right
]/D%\right
).
$
% \label{cond1a}
%\end{equation}
Discussing separately the cases $\alpha_3>,< 1/2$ and recombining results, we obtain the range for possible structure formation
\be
 2 > \alpha_3 >1:&& {1\over 4 \alpha_3}\left (1-{1\over (2 \alpha_3-1)^2}\right
 ) <\alpha_1<\frac{2}{9\alpha_3}\nonumber\\
 1 > \alpha_3 >0:&& 0<\alpha_1<\frac{2}{9\alpha_3}
 \label{reg}
\ee
represented in Fig. \ref{InstabReg} as a phase diagram for the
freezing and structure parameters.

The structure parameters 
$\alpha_1$ determines the brine
channel formation. A small $\alpha_1$ means low temperatures or low
salinities and consequently a freezing process
with a uniform ice phase for sufficiently large $\alpha_3$ and a precipitate of salt. In contrast at higher $\alpha_1$ there are higher
temperatures or higher salinities inducing a melting with
a uniform liquid water phase and dissolved salt. The spatial
structures can only appear in the instability region
which starts at the maximal point $\alpha_1 = 1/9$ at $\alpha_3 = 2$.
The description of the instability region does
not involve a restriction on the diffusivities of salt and water.
This is different from the model of \cite{KMG09}, which
describes structure formation in sea-ice in terms of Turing structures.

\section{Determination of parameters} 

Before solving (\ref{f2}) numerically we use (\ref{aa0}) to determine the values
of $\alpha_1$ and $\alpha_3$ in terms of water properties.

%with the super-cooling temperature $T^0_c=-40^0C$ (\cite{Nev,Dor}). 
Using the latent heat of the phase
transition from water to ice $\Delta H = 6 kJ/
mol$  and a dissociation ratio of $x =
(n_{Na^+} + n_{Cl^-})/n_{H_2O} = 1/50$, the
Clausius-Clapeyron relation yields a freezing point depression of
$
\Delta T_{cc} = -
\frac{xRT^2}{\Delta H}=-2K$
in agreement with the natural value of $\Delta T = -1.9$ K. 
After
a super-heating of more than 5$^\circ$C, homogeneous nucleation occurs in the metastable state \cite{Bau84}. For fresh water ($T_c = T_0 = 273.15$K and $T^0_{c,s} = T^0_c = 233.15$K) from
equation (\ref{45}) follows that $T_1 = 278.11$K ($4.96^\circ$C) as the upper limit of super-heating in agreement with the experiment \cite{Bau84}.
According to (\ref{45}) and (\ref{16}) and (\ref{inst}) these super-heating and freezing temperatures are realized by choosing $\alpha_1=0.2$ and $\alpha_3=0.9$
 The structure
parameter $\alpha_3 = 0.9$ leads to a  freezing point temperature of
$-1.9^\circ$C ($T_c = 271.25$K) for seawater of salinity $35$g/kg ($\rho_0 = 0.6$mol $NaCl/53$ mol $H_2O
= 0.0113$) and represents therefore a realistic description of super-cooling pure water.

%Our choice of the freezing parameter
%$\alpha_1 = 0.2$ represents a temperature $T_2 = -8.2^\circ$C.
% in agreement with a transition temperature of mirabilite. 
Furthermore, the specific heat $c$ is dependent on $\alpha_3$ as
\be
c&=&\left . \!-T\frac{\partial^2f(\psi_0^+(T))}{\partial
  T^2}\right |_{T=T^0_{c}}
\nonumber\\
&=&\frac{\tilde{\alpha}^2_1 T^0_c}{2\alpha_3}\!\left(\!1\!+\!\frac{3}{\sqrt{1\!+\!36\tilde{\alpha}_1\alpha_3(T^0_c\!-\!T)}}
\!\right)
\nonumber\\
&=&\frac{4}{81}\frac{T^0_c}{\alpha^3_3(T_c\!-\!T^0_c)^2}.\label{c1}
\ee
We set the energy scale to be the difference of
the latent heat of water freezing $K_E=L(0^\circ C)-L(-40^\circ C)=98 {J/g}$ \cite{Nev}.
The resulting specific heat in our theory yields $c_{spec}=K_E c=2.14J/gK$ 
which compares well with the
experimental value of $c_{exp}=2J/gK$. This shows that the choice of the 
structure
parameter $\alpha_3=0.9$ is in agreement with the specific heat too.

The parameters $\alpha_1$ and $\alpha_3$ define the local portion of the free energy in
a system with uniform order parameter and salinity. The spatial inhomogeneity of the system
is described by the third parameter of the model $D = D_{\rm ice}/D_{\rm salt}$. At the freezing temperature of seawater of  $-1.9^\circ$C, the study in \cite{Maus} predicts $D_{salt,-1.9^\circ C} = 0.62 \times 10^{-5}cm^2/s$. The $D_{\rm ice}$ can be linked to the reorientation rate of the $H_2O$-molecules and the correlation length 
%\cite{Thoms} 
which leads with realistic numbers \cite{Bo97,Eis} to $D_{\rm ice} = 0.33 \times 10^{-5}cm^2/s$ and finally to a ratio $D_{ice}/D_{salt} = 0.47$.

\begin{figure}[h]
\centerline{
\parbox{4.5cm}{ 
\includegraphics[width=4.5cm]{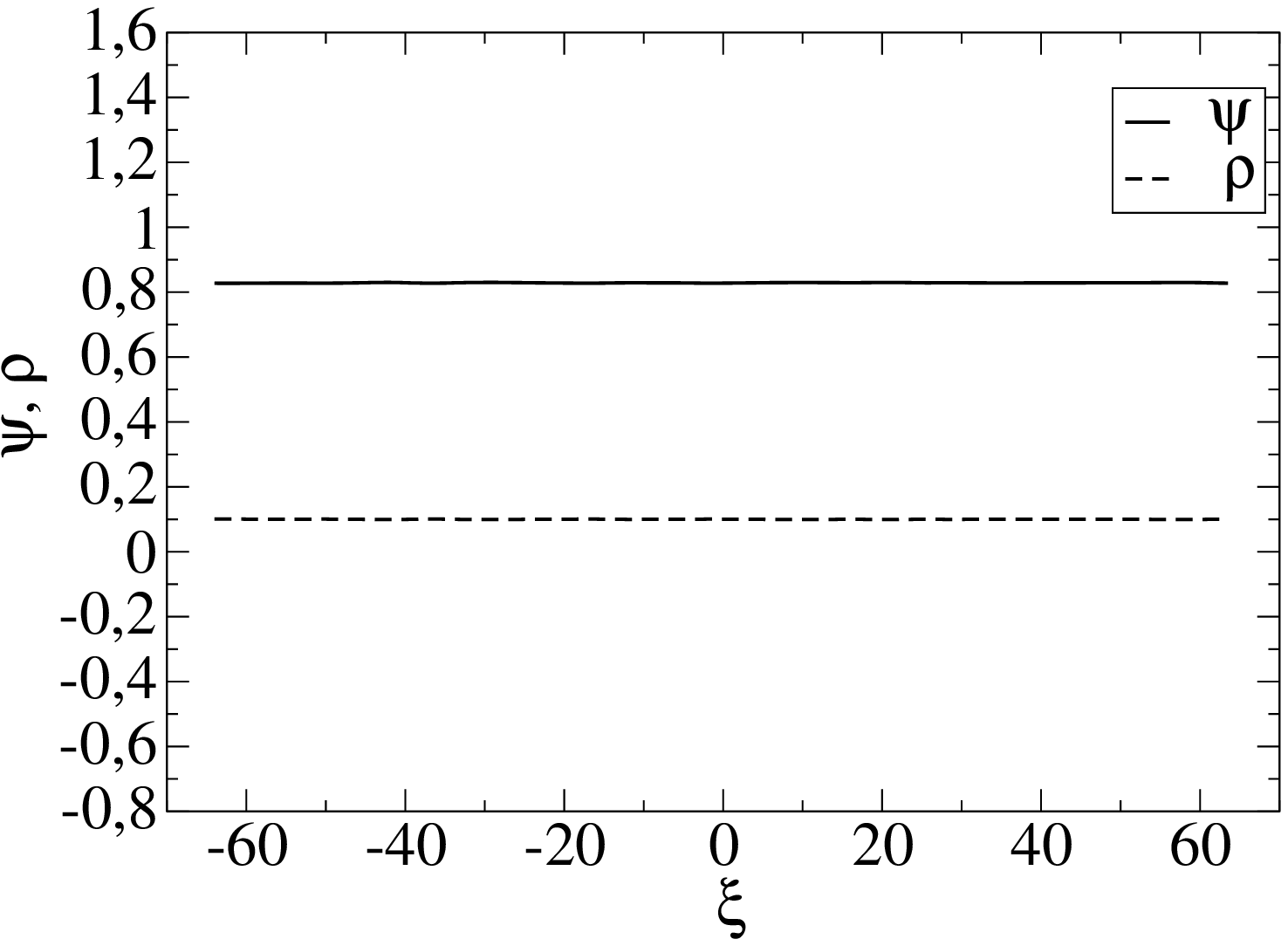}
\includegraphics[width=4.5cm]{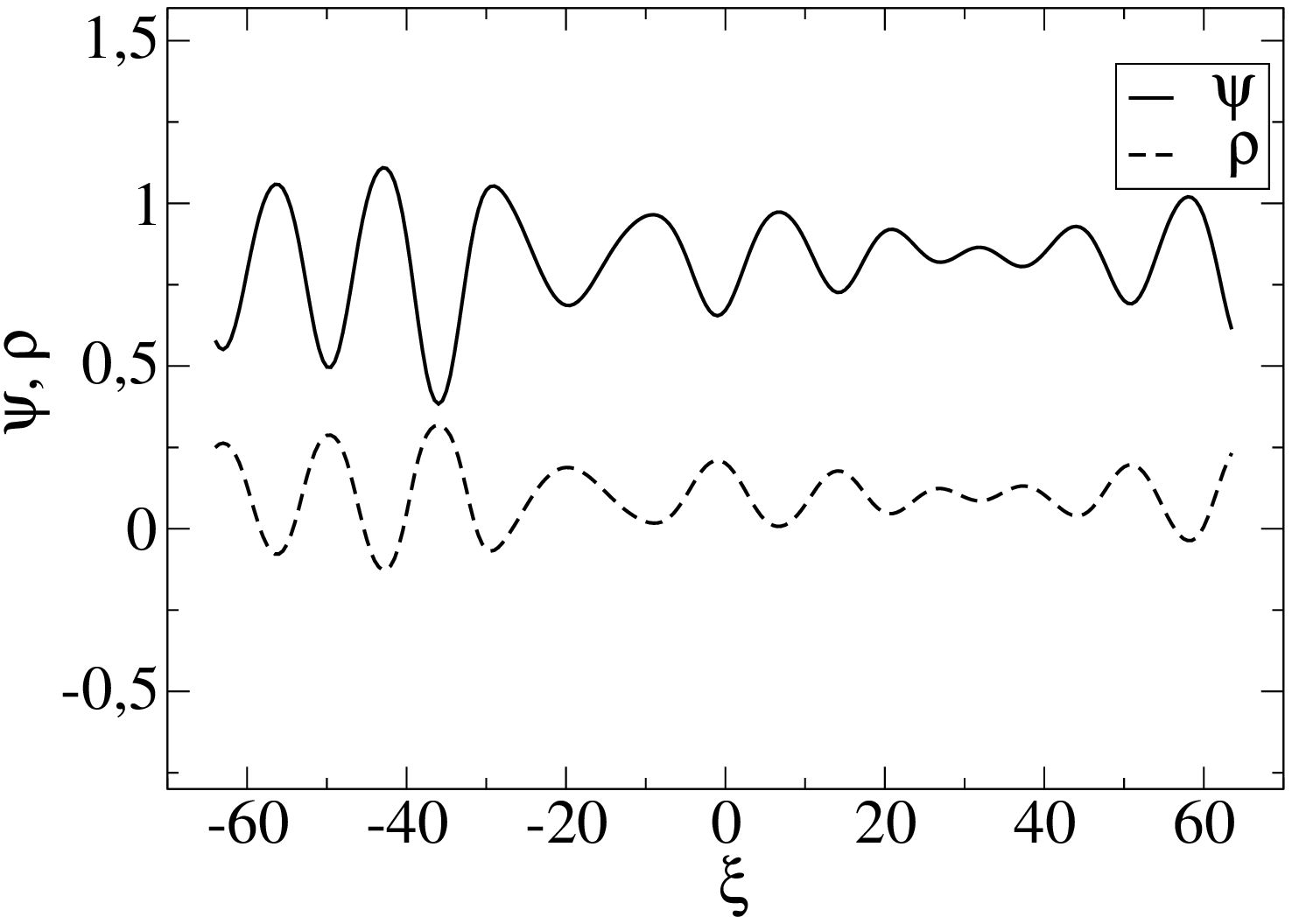}
\includegraphics[width=4.5cm]{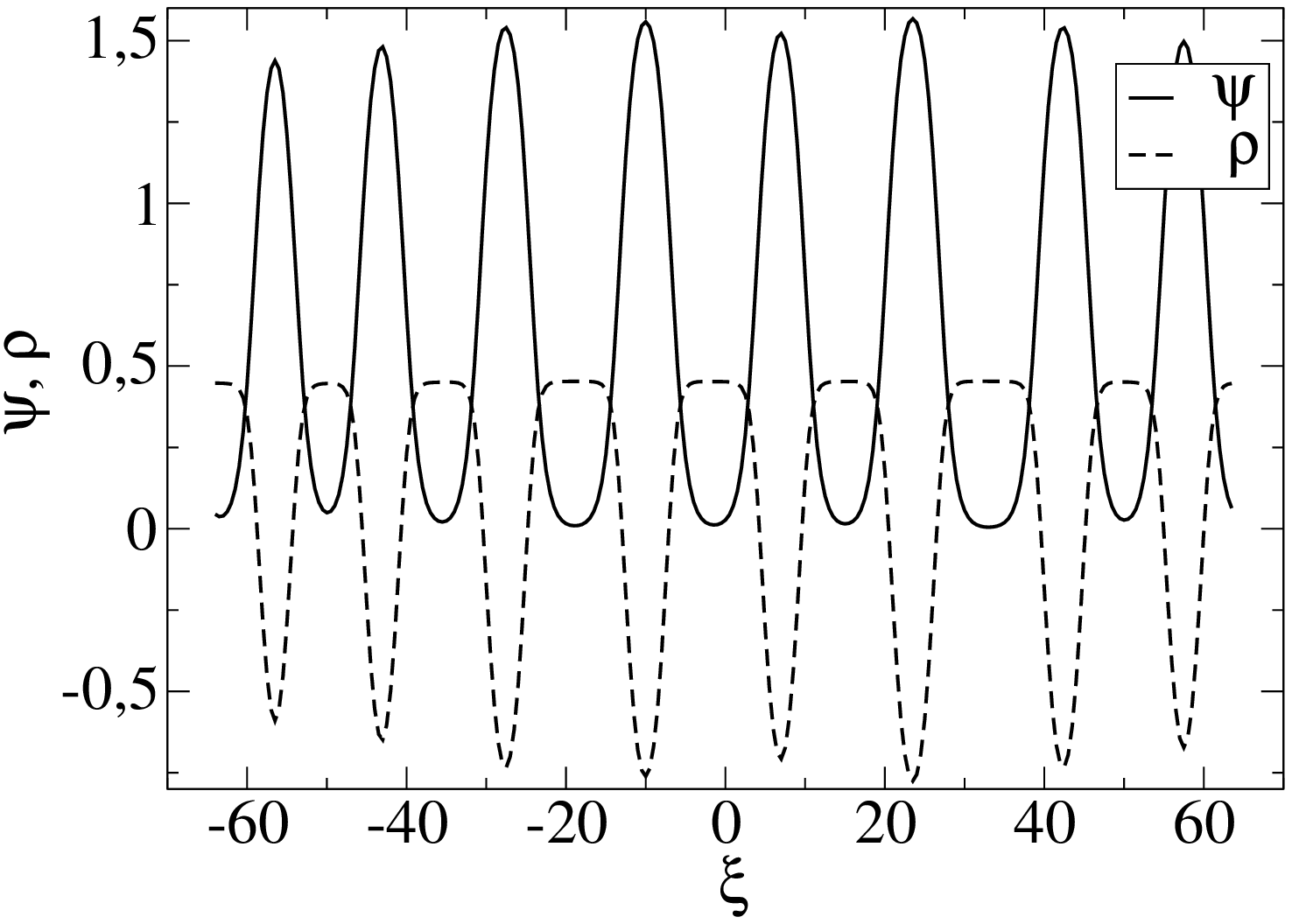}
}
\parbox{4cm}{
 \includegraphics[width=4cm,angle=0]{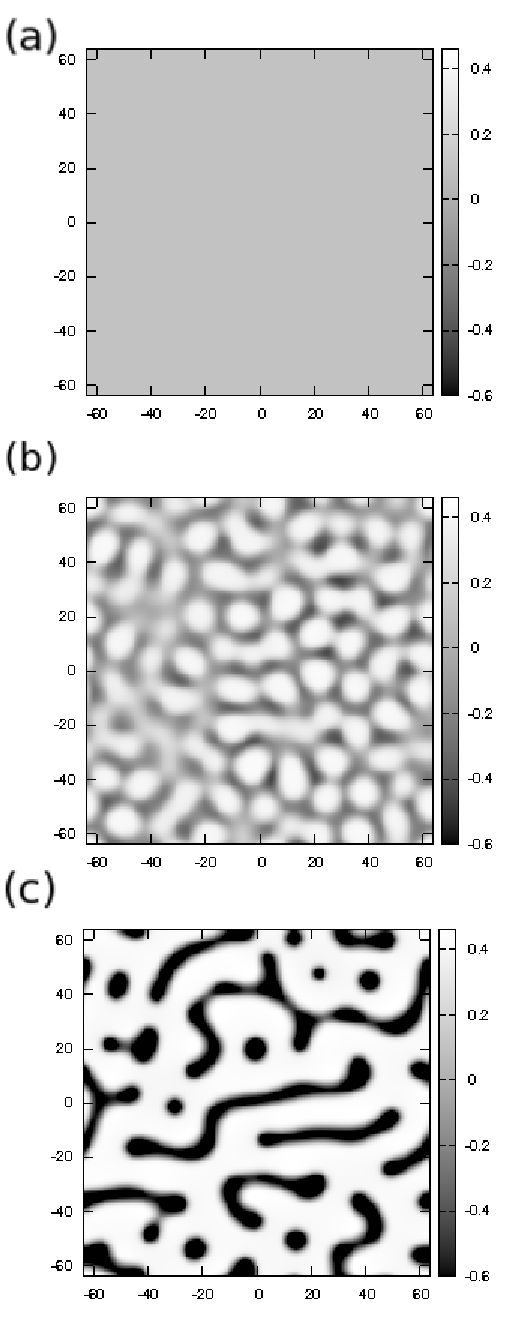}
}
}
  \caption{Time evolution of the order parameter
  $\psi$ and salinity $\rho$ as deviation from mean salinity for 1D (left) and the salinity for 2D (right) 
versus spatial 
  coordinates for $\tau=10, 150, 500$ (from above to below)
  with the initial random distribution $\psi(\tau=0)=0.9$
  and $\rho(\tau=0)=0.1 \pm 0.001 N(0,1)$. The parameters
  are $\alpha_3 = 0.9$, $\alpha_1 = 0.1$, and $D = 0.5$.}
\label{Num2D}
\end{figure}

\section{Time evolution and pattern formation}

Now we integrate the equation system (\ref{f2}) numerically in one and two space 
dimensions by an exponential time
differencing scheme of second order (ETD2) \cite{CoMa}. 
We have a stiff
differential equation of the type $\dot y=r y+z(y,t)$ with a linear term $r
y$ and a nonlinear part $z(y,t)$.  
The linear equation is solved analytically and the integral over the nonlinear
part is approximated by a proper finite differencing scheme.

The evolution of the order parameter $\psi$ 
and the salinity $\rho$ in one and two dimensions is shown in Fig. \ref{Num2D}. 
The quantities $\psi$ and $\rho$ are complementary in phase. Due to the second equation of (\ref{f2}), the conservation of salinity $\int dx \rho(t,x)=c$ is ensured. We can absorb this mean salinity into $\rho\to \rho-c$ leading to a mere shift in $\alpha_1'-c+\rho_0=\alpha_1$ which means we consider with $\rho$ the deviations from a mean salinity $c$ and the total salinity remains positive.  
Regions of high salinity correspond to the water phase and regions 
of low salinity correspond to ice domains. We see that one single mode develops given by the wave number $\kappa_c$. Similar to the one-dimensional
case, we see the formation of one dominant wavelength also in two 
dimensions. this can be understood as the maximum of unstable wavelengths (\ref{rootslam}) which becomes
\be
\kappa^2_c  &=& {\psi_0\over (D\!-\!1)^2}\biggl [ {(D\!-\!1) \left(1\!-\!2 \alpha_3 \psi_0\right)
                \!-\! 2 \psi_0}
\nonumber   \\&+& 
{(D\!+\!1)\psi_0^{1/2} \sqrt{(D\!-\!1) (2 \alpha_3
               \psi_0\!-\!1)\!+\!\psi_0}\over \sqrt{D}}\biggr ].
\label{kcrit}
\ee
The critical wave number sets the length scale on which phase 
separation occurs and is visible as the dominating coarse graining mode in figure \ref{Num2D}.
The size of solidification structures depends on the super-cooling relative to the
freezing temperature $T_c$. The higher the super-cooling, the more rapidly water freezes and the
smaller the structures become.

\begin{figure}[ht]
\centerline{\parbox{8.5cm}{
\parbox{4cm}{
\includegraphics[width=4cm,angle=0]{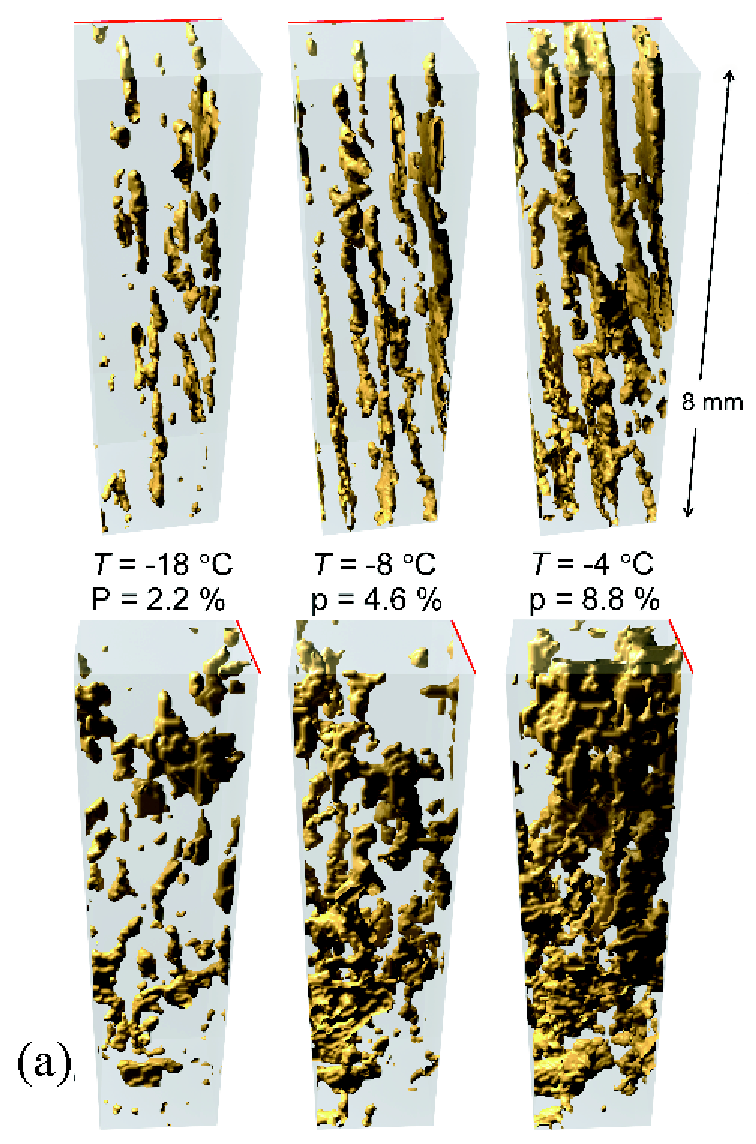}\\
(b)\includegraphics[width=3.2cm,angle=0]{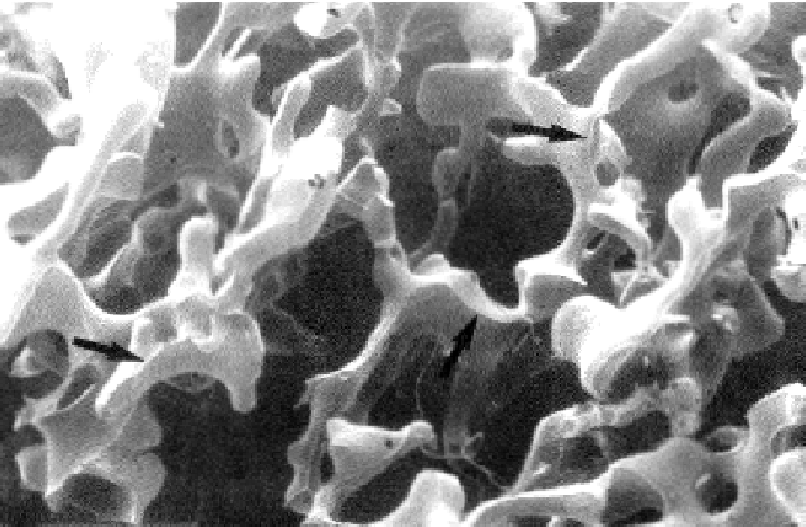}
}
\parbox{4.3cm}{ 
(c)\includegraphics[width=3.7cm,height=3.7cm,angle=0]{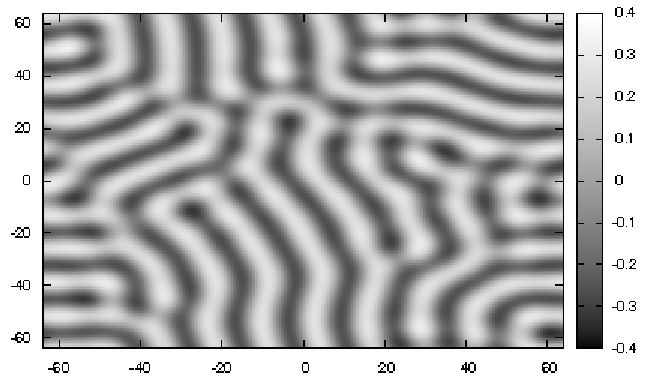}\\
(d)\includegraphics[width=3.8cm,angle=0]{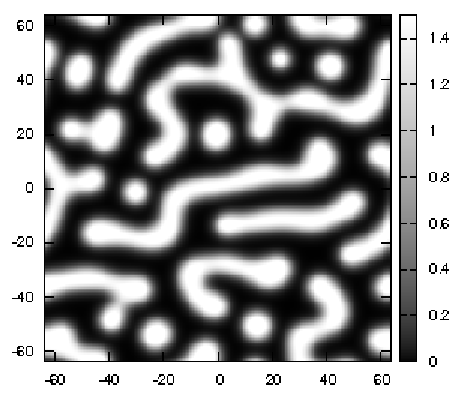}
}}
}
\caption{\label{comp} (a) Imaging brine pore space with X-ray computed tomography (image from \cite{Pringle}). The
upper images shows the view approximately along the brine layers. The view across the brine
layers is shown in the bottom images.
(b) Scanning electron microscopy image of a cast of brine
  channels (entrapment) \cite{Wei}, (c) Turing structure after long time 
%for $\tau=400$
  \cite{KMG09}, (c) long-time phase-field structure %for $\tau=500$ 
from figure \ref{Num2D}.}
\end{figure}

\section{Comparison with experiments}

Concerning the experiments we suggest three types of comparisons: (i) morphology, (ii) per\-col\-ation thre\-shold and (iii) structure size where our model describes realistic parameters. We will start with the morphology.
For the web of brine channels one observes different textures for instance
granular ice, columnar-granular structures or plate ice.
Fig. \ref{comp}(b) shows a measurement yielding 
granular texture \cite{Wei} without prevalent
orientation. 
In figure \ref{comp}(c) we have chosen the best fit of the former Turing-model \cite{KMG09} to the structure size.
If we compare with the simulation of our phase-field model in Fig. \ref{comp}(d), the texture of
the cast of brine channels seems to be better described by our present model than by the Turing model. Though the absolute size is not so much different, the three parameters of the Turing model had been adjusted to fit the structure as best as possible. Here, with the phase-field model, we have chosen parameters according to the thermodynamic properties of water and have obtained the structure as a consequence of these parameters.

The structure
of the brine pore space of single crystals \cite{Pringle} is shown in the iso-surface plots in Fig. \ref{comp}(a). The upper images clearly
show near-parallel intra-crystalline brine layers. The view across the layers (bottom
images) show brine layer textures much more complicated than suggested by the simple
model of parallel ice lamellae and parallel brine sheets illustrated in Fig. \ref{comp}(b). 
Depending on the temperature, the images show a brine pore porosity from $p=2.2\%-8.8\%$. The connectivity 
increases with porosity as the pore space changes from isolated 
brine inclusions at $p = 2.2\%$ to extended, near-parallel layers 
at $p = 8.8\%$. The thermal 
evolution of the brine pore space with percolation theory was characterized in \cite{Pring09} where a 
connectivity threshold was found at a critical volume fraction $p_c = 4.6\%$. 
Below $p_c$ there are no percolating pathways spanning through the sample, i.e.
the brine is trapped within the intra-crystalline brine layers. 

\begin{figure}[]
\subfigure[\ Histogram of figure \ref{comp}a along the brine layers (above) and across the layer with increasing temperature from $-18,-8,-4^\circ$C (below).]{\includegraphics[width=8.5cm,angle=0]{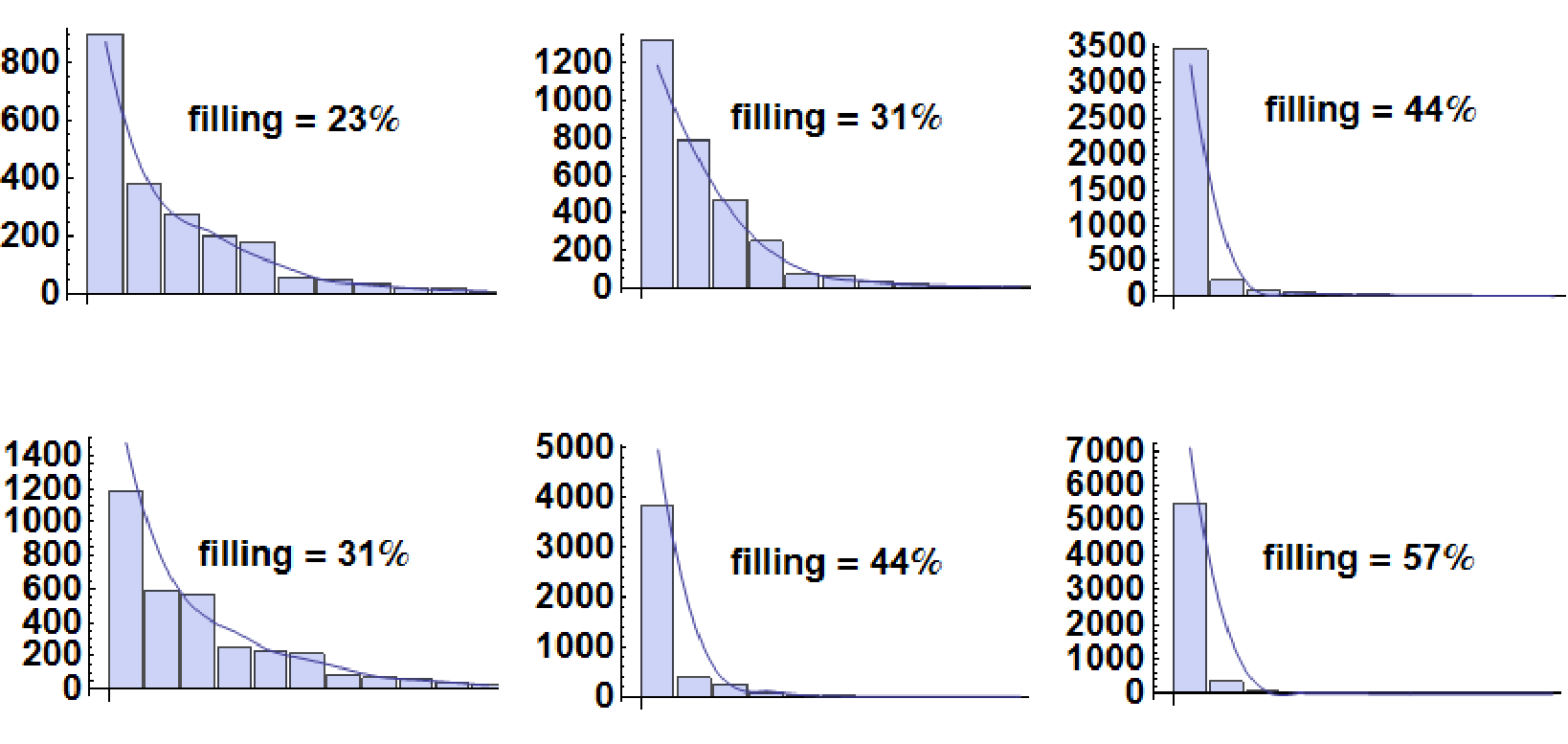}}

%\hspace*{0.5cm}
\subfigure[\ Histogram of the numerical result in figure \ref{Num2D} 
with $\alpha_1' = 0.1$, $D = 0.5$, and $\alpha_3 = 0.9$ (left) compared to $\alpha_3 = 1$ (right).]{\includegraphics[width=8.5cm,angle=0]{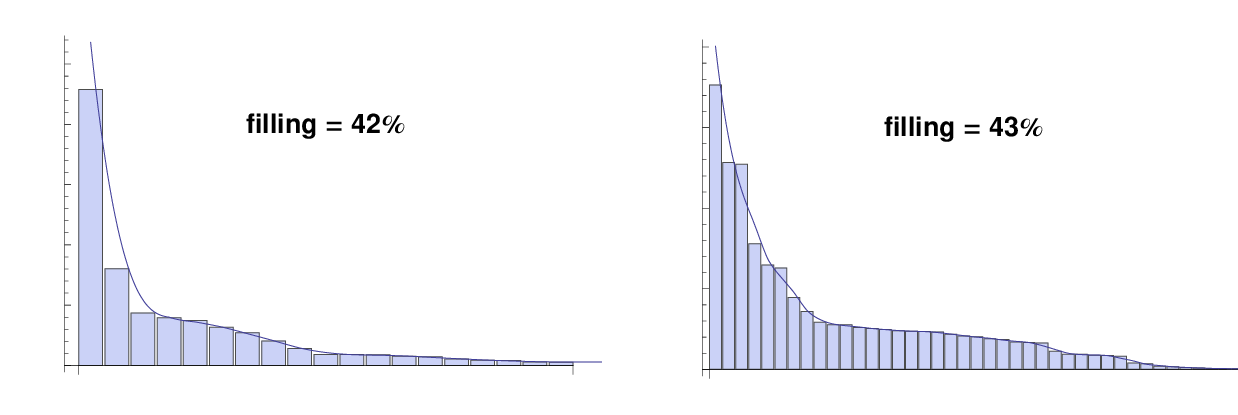}}
\caption{\label{pringle1} 
The histograms of connected clusters.} 
\end{figure}

Lets quantify this statement by a cluster-size analysis of the figures \ref{comp}a where the corresponding histograms are given in figure \ref{pringle1}. As one finds, the percolation transition is visible around 44\% filling in the range of ($-8$$^\circ$C, $-4$$^\circ$C). Now we compare with our simulation varying the parameter $\alpha_3=0.9$ and $\alpha_3=1$. We see that the same percolation threshold appears with a comparable histogram for $\alpha_3=0.9$. This shows that the parameters of our model which where chosen to reproduce the thermodynamics, allows also to describe realistic morphological structures.

%\noindent\parbox[]{13.5cm}{
%\parbox[]{7.5cm}{
%}
%\parbox[]{6cm}{
%\begin{figure}[]
%\includegraphics[width=6cm,angle=0]{own1.eps}
%\caption{\label{pringle1a} 
%The histogram of connected clusters of the numerical result in figure \ref{Num2D} 
%with $\alpha_1' = 0.1$, $D = 0.5$, and $\alpha_3 = 0.9$ (left) compared to $\alpha_3 = 1$ (right).}
%\end{figure}
%}
%}

Next we compare the size of the obtained structure with the size of pure sea
ice platelets \cite{W92,Pringle,Gol2} which separate regions of concentrated seawater. The fastest-growing wave-vector
$\kappa_c(D,\alpha_1,\alpha_3)$
%\be
%\kappa^2_c & =  {\psi_0^+\over (D-1)^2}\biggl [ {(D-1) \left(1-2 \alpha_3 \psi_0^+\right)
%                - 2 \psi_0^+}\nonumber   \\
%           &+  {(D+1)\sqrt{\psi_0^+} \sqrt{(D-1) (2 \alpha_3
%               \psi_0^+-1)+\psi_0^+}\over \sqrt{D}}\biggr ]
%\label{kcrit}
%\ee
sets the length scale on which phase 
separation occurs. The size of the structure can be estimated
by $2\pi/\kappa_c$. 
With the help of (\ref{dT}) and remembering the dimensionless values introduced before (\ref{f2}), the critical domain size of the phase-field structure
as a function of the freezing point depression takes the value
\be
\lambda_c ={2\pi \over k_c}
=
{2\pi\over \kappa_c}{
h\over a_2}
=
{2\pi\over \kappa_c}
\sqrt{
D_{salt}\rho_0\over 
\tilde a_1|\Delta T|}
\label{47}
\ee
where one gets with the parameters $\alpha_3=0.9, \alpha_1=0.2$ and $D=D_{ice}/D_{salt}=0.5$ a dimensionless pattern size of $13.81$. Our choice of the freezing parameter
$\alpha_1 = 0.2$ represents a super-cooling $\Delta T_{sup} = 6.3$K. The rate of reorientation of the $H_2O$-molecules determines
$\tilde a_1 = 1/[(T_0 - T^0_c)\tau_d(T_0)] = 1250 K^{-1}s^{-1}$. With these parameters we obtain from (\ref{47}) a critical domain size $\lambda_c = 0.8\mu$m in agreement with the sea ice platelet spacing
$\lambda_{max} \approx 1\mu$m obtained from morphological stability analysis \cite{W92} or percolation theory \cite{Gol2,Pringle}.

We consider now the size of phase-field structures for natural conditions which is
given by the upper limit of the instability region shown in Fig. \ref{InstabReg}. With a structure
parameter $\alpha_3 = 1.99$ and a freezing parameter $\alpha_1 = 0.111482$ one has a realistic description
of seawater at $0.032$K super-cooling and a lower limit of the super-cooling region of fresh
water at $-18.78^\circ$C. For this growth condition we obtain $2\pi/\kappa_c = 4975.25$ as dimensionless
structure size  and using equation (\ref{47}) the critical domain size is $\lambda_c =198\mu$m in agreement with the observed values. Brine inclusions \cite{Wei} have scales from $3-1000\mu$m, where the average dimensions is typically $200\mu$m. 

\section{Summary}

To summarize, a model for the formation of salty water channels (brine entrapment) in sea ice has been developed which consists of two coupled order parameters, the tetrahedricity and the
salinity preserving the mass conservation of salinity. 
The linear stability analysis provides a phase diagram in terms of two model parameters
indicating the region where spatial structures can be formed due to the instability of the
uniformly ordered phase. The region of instability is determined exclusively by the freezing parameter and the (specific heat) structure parameter and not by the diffusivity as it was the case in the reaction-diffusion Turing model \cite{KMG09}. This allows to link these model parameters to thermodynamical properties of water like super-heating, super-cooling, freezing temperature and specific heat simultaneously. 

With the help of these model parameters we solve
the time-dependent coupled evolution equations and find a brine channel
texture in agreement with the experimental values. That the physical justification of the parameters by other properties of water leads here to a better description of the brine channel texture, we attribute to the mass conservation invoked in the present model. 

The presented model does not include yet the heat transfer. We have merely concentrated on structure formation at short-time scales to consider the processes adiabatically. Therefore the model should be extended to include the temperature field as a third order parameter. The inclusion of a velocity field is also necessary to describe real situations since convective motions certainly are expected to be present.

\begin{acknowledgements}
This work was supported by DFG - priority program SFB 1158. The financial support by the Brazilian Ministry of Science 
and Technology is acknowledged.

\end{acknowledgements}
The model set-up and linear stability analysis has been performed by all authors. The relation of model parameters to properties of water has been derived by S. Thoms and B. Kutschan. Picture analysis and histograms have been provided by K. Morawetz. Numerical codes were performed by B. Kutschan.
%The model has been set up by S. Thoms, the linear stability analysis and histograms have been formulated by K. Morawetz as well as the text and the numerical codes were performed by B. Kutschan. The relation of model parameters to properties of water has been developed by B. Kutschan and S. Thoms.
\bibliographystyle{spphys}       % APS-like style for physics
\bibliography{seaice,afp}   % name your BibTeX data base

%\begin{thebibliography}{}
% etc
%\end{thebibliography}

\end{document}